\documentstyle[12pt]{article}
\begin{document}
\def\thefootnote{\fnsymbol{footnote}}
\vspace{ .7cm}
\vspace*{2cm}
\begin{center}
{\LARGE\bf Neutrino oscillations in a $3\nu_L+3\nu_R$ framework
with five light neutrinos }\\
\vspace{1 cm}
{\Large  Daijiro Suematsu}
\footnote[3]{Email address: suematsu@hep.s.kanazawa-u.ac.jp}
\vspace {1cm}\\
{\it Department of Physics, Kanazawa University,\\
        Kanazawa 920-1192, Japan}
\end{center}
\vspace{2cm}
{\Large\bf Abstract}\\  
We propose a neutrino mass matrix model in which five neutrino 
species remain light through the seesaw mechanism within a 
supersymmetric $3\nu_L+3\nu_R$ framework.
We construct such a model based on the nonrenormalizable terms in 
the superpotential constrained by the discrete symmetry 
which may be expected in the models at the high energy scale such as
superstring.
We study the possible oscillation phenomena by fixing mass parameters 
so as to explain the solar and atmospheric neutrino deficits 
and also include a candidate of the suitable dark matter. 
We also discuss the charged lepton mass matrix based on this neutrino 
model. LSND results may be consistently explained within this model.
\newpage
\setcounter{footnote}{0}
\def\thefootnote{\arabic{footnote}}
\section{Introduction}
Neutrino mass is one of unsolved problems in the present particle physics,
although it is a very important issue which can be a clue to go beyond the
standard model.
Experimentally, in both particle physics and astrophysics
there are a lot of indications for massive neutrinos \cite{rev}.
Analyses of observations of the solar neutrino \cite{solar} and the 
atmospheric neutrino \cite{atm} 
suggest the existence of neutrino oscillations.
Especially, a recent SuperKamiokande observation of the Zenith angle
dependence of the atmospheric $\nu_\mu$ strongly suggests that $\nu_\mu$
oscillates to $\nu_\tau$ \cite{neut98}.
An interesting feature in these indications is that they may require 
wide range mixing angles, in particular, a maximal mixing 
among different neutrinos in addition to hierarchically small mass
eigenvalues or closely degenerating mass eigenvalues. 
Although this smallness of masses may be considered to be 
explained by the seesaw mechanism \cite{seesaw} in general, 
the hierarchy of masses and the mixing structure will be completely 
dependent on models.

On the other hand, it has also been suggested that 
neutrinos with suitable masses could be a dark matter candidate for the
explanation of astrophysical observations related to 
the structure formation of the universe.
As such examples, we may list an active neutrino with 
${\cal O}(10)$ eV mass in
the hot dark matter scenario (HDM) \cite{bes}, 
a sterile neutrino with ${\cal O}(10$-$10^2$) eV mass 
in the warm dark matter scenario (WDM) \cite{dw} 
and neutrinos with ${\cal O}$(1) eV mass in 
the cold + hot dark matter scenario (CHDM) \cite{dark,phkc}. 
When we consider the neutrino models, we may need accomodate this
feature to the models. If we impose 
such a requirement, we can restrict the models in a rather severe way.

Now various experiments using reactors, accelerators and underground 
facilities are proceeding and planned. In the near future these
experimental results will be presented to inform us details of the
neutrino sector and then 
the predictions for various neutrino phenomena including oscillations 
on the basis of possible neutrino models will also be very useful.
Under this situation it seems to be a worthy and interesting subject
to consider various types of model which can explain both of 
these neutrino deficits 
consistently and present a dark matter candidate.

The introduction of a sterile neutrino is one way to this direction
and a lot of works have been done by now \cite{pmodel,sat,sue1,sue2}.
It has also been suggested that sterile neutrinos may play important
roles in various phenomena \cite{ster,asym1,ster2}.
Although the existence of sterile neutrinos is an interesting
possibility, it seems to be rather difficult to
find its natural candidate in particle physics models.\footnote{
For example, a candidate other than the right-handed neutrino 
has been proposed in Ref. \cite{modul}.}
One of such reasons is that it is not so easy to introduce a 
substantial mixing between active neutrinos and light 
sterile neutrinos in the natural way \cite{lang}.

The aim of this paper is to propose such a candidate and
analyze neutrino oscillation phenomena on the basis of it.
In this study the neutrino sector is extended into
$3\nu_L+3\nu_R$ species, among which only five neutrinos remain 
light through the seesaw mechanism based on the mixings with a heavy
right-handed neutrino $(N_R)$.
Two light right-handed neutrinos behave as sterile neutrinos.
This seems to be natural from a viewpoint of the generation
structure of other quarks and leptons, although the seesaw mechanism
works in a different way from the one in the ordinary grand unified models 
such as SO(10). 
To realize the substantial mixing between these sterile neutrinos and
active neutrinos we will consider the nonrenormalizable interactions
coming from the fundamental theory in a high energy region. 
This may be considered as an example of the scheme proposed in Ref. 
\cite{lang}.  
Phenomenologically, as partly discussed in Refs. \cite{sue1,sue2},
this model can give a framework to explain both deficits of the solar and
atmospheric neutrinos consistently and also to present a candidate 
of the dark matter with a suitable mass for a certain kind of dark 
matter scenario.

This paper is organized as follows.
In section 2 we give a brief review on the oscillation parameters to
fix our notation.
In section 3 we introduce our model and discuss its theoretical
background. A possibility to realize this model in the supersymmetric
framework in the basis of
nonrenormalizable terms in the superpotential is studied.
Its detailed phenomenological analyses are presented in section 4. We show
two typical parameter settings which have different features.
Other oscillation processes than the ones related to the solar and
atmospheric neutrino deficits are also discussed in each case.
LSND results reported in Ref. \cite{lsnd} may be simultaneously 
explained in this model.
Section 5 is devoted to the summary.

\section{Oscillation parameters}
At first we briefly review a basic formula for the neutrino
oscillation.
We define a mixing matrix $V$ as 
$\nu_f=\sum_\alpha V^\ast_{\alpha f}\tilde\nu_\alpha$ 
where $(V^\dagger)_{f\alpha}=V^\ast_{\alpha f}$ and 
$\tilde\nu_\alpha~(\alpha=\bar 1, \bar 2, \bar 3,\cdots)$
is a mass eigenstate. $\nu_f~(f=e, \mu, \tau, \cdots)$ is a 
weak eigenstate chosen in a way
that both leptonic charged currents and a charged lepton mass matrix
are diagonal. Thus $V$ can be written as
$V=U^{(\nu)}U^{(l)\dagger}$ by using diagonalization matrices 
$U^{(l)}$ and $U^{(\nu)}$ of the charged lepton and neutrino mass 
matrices $M^{(l)}$ and $M^{(\nu)}$,
\begin{equation}
M_{\rm diag}^{(l)2}=U^{(l)}M^{(l)\dagger}M^{(l)}U^{(l)\dagger}, \qquad
M_{\rm diag}^{(\nu)}=U^{(\nu)}M^{(\nu)}U^{(\nu)T}.
\end{equation}
In the following discussion, $V$ is assumed to be real, for simplicity.
Then $V$ satisfies an orthogonality condition\footnote{
We neglect CP phase here.
In a model studied in this paper there are three charged lepton
flavors and five light neutrino flavors. When we define $V$, 
it is necessary to extend $U^{(l)}$ into a $5\times 5$ matrix by adding 
formally 1's as diagonal elements.}:
\begin{equation}
\sum_f V_{\alpha f}V_{\beta f}=\delta_{\alpha\beta}, \qquad
\sum_\alpha V_{\alpha f}V_{\beta f}=\delta_{\alpha\beta}.
\end{equation} 

Using mass eigenstates, a time evolution equation of neutrinos in
a vacuum is given as
\begin{equation} 
i{d\over dt}\tilde\nu_\alpha ={M_\alpha^2\over 2E}\tilde\nu_\alpha,
\end{equation}
where $E$ is energy of neutrinos and $M_\alpha$ stands for an {\it $\alpha$-th}
neutrino mass eigenvalue.
This equation can be easily solved as
\begin{equation}
\tilde\nu_\alpha(t)=\exp\left(-i{M_\alpha^2\over 2E}t\right)\tilde
\nu_\alpha(0).
\end{equation}
By transforming this into a solution in terms of 
the weak eigenstates, the transition
probability for $\nu_f \rightarrow
\nu_{f^\prime}$ during the time interval $t$ is expressed as 
\begin{eqnarray}
&&P_{\nu_f\rightarrow\nu_{f^\prime}}(t)=\vert\langle\nu_{f^\prime}(t)
\vert\nu_f(0)\rangle\vert^2  \nonumber \\ 
&&\hspace*{1cm}=\sum_\alpha V_{\alpha f^\prime}^{2}V_{\alpha f}^{2}
+2\sum_\alpha\sum_{\beta(>\alpha)}
V_{\alpha f^\prime}V_{\alpha f}V_{\beta f^\prime}V_{\beta f}
\cos\left( {\Delta M^2_{\alpha\beta} \over 2E}t\right),
\end{eqnarray}
where $\Delta M^2_{\alpha\beta}\equiv \vert M_\alpha^2-M_\beta^2\vert$.
If we use a following relation derived from the orthogonality of $V$:
\begin{equation}
\sum_\alpha V_{\alpha f^\prime}^{2}V_{\alpha f}^{2} 
=-2\sum_\alpha\sum_{\beta (>\alpha)}
V_{\alpha f^\prime}V_{\alpha f}V_{\beta f^\prime}V_{\beta f},
\end{equation}
we can obtain
\begin{equation}
P_{\nu_f\rightarrow\nu_{f^\prime}}(t)=-4\sum_\alpha\sum_{\beta(>\alpha)}
V_{\alpha f^\prime}V_{\alpha f}V_{\beta f^\prime}V_{\beta f}
\sin^2\left( {\Delta M^2_{\alpha\beta} \over 4E}t\right).
\end{equation}

Here we note that a contribution from a $\alpha\beta$-sector 
to the $\nu_f \leftrightarrow \nu_{f^\prime}$ oscillation is represented by
parameters $(\Delta M_{\alpha\beta}^2,
~-4V_{\alpha f^\prime}V_{\alpha f}V_{\beta f^\prime}V_{\beta f})$.
In a sector where $V_{\alpha f}$ and $V_{\beta f^\prime}$ correspond to the 
diagonal elements the amplitude ( or the mixing
factor ) reduces to $-4V_{\alpha f^\prime}V_{\beta f}$, 
as long as the mixings 
with other neutrino flavors are 
sufficiently small and $V_{\alpha f}\sim V_{\beta f^\prime}\sim 1$ 
is satisfied. In such a case  
these parameters can be understood as the usual two flavor
oscillation parameters $(\Delta m^2, ~\sin^22\theta)$.
On the other hand, if the mixings with other flavors are not so small
 in the sectors specified by the off-diagonal elements $V_{\alpha f}$
 and $V_{\beta f^\prime}$, there appear 
new contributions which are induced due to the existence of many
neutrino flavors.
They may be understood in such a way that an
additional mixing between $\nu_{f_1}$ and $\nu_{f_2}$ induces the  
$\nu_f\leftrightarrow\nu_{f^\prime}$ oscillation through the flavor
mixings $\nu_f-\nu_{f_1}$ and $\nu_{f^\prime}-\nu_{f_2}$.
Thus these processes are generally recognized as higher order effects 
in comparison with the direct two flavor oscillation 
concerning the mixing factors.  
Anyway it is necessary to be careful when we apply Eq. (7) to the results
on the oscillation parameters obtained by the two flavor 
analysis of the solar and atmospheric
neutrino problems. In particular, matter effects on the oscillations
will not be observed in a simple way through an analytical study 
in the cases with many neutrino flavors.

\section{A model with sterile and active neutrino mixings}
\subsection{Basic framework}
We consider a model containing three generation left-handed neutrinos
$\nu_{f_L}~(f_L=e,\mu,\tau)$, and right-handed neutrinos 
$N_{f_R}~(f_R=A,B,C)$.
As a guiding principle to construct a neutrino mass matrix in this
kind of models phenomenologically and reduce the free parameters
systematically, we take a viewpoint that the realization 
of the mixing angle required
in the solar and atmospheric neutrino problems is the most important
clue. 
Along this line we first prepare mass terms which can produce such a mixing
structure among the five light neutrinos through the seesaw mechanism.
After this procedure we introduce mass correction to resolve the mass
degeneracy in a consistent way with this mixing structure.

Following this strategy, 
we require that six neutrino species have mass terms which are written as
\begin{equation}
-{\cal L_{\rm mass}}=\sum_{f,f^\prime =1}^5m_{ff^\prime}\psi_{f}\psi_{f^\prime}
+\sum_{f=1}^5m_{f}\psi_{f}\bar N_{C}
+{1 \over 2}M\bar N_{C}\bar N_{C}+{\rm h.c.},
\end{equation}  
where $\psi_f$ represents $\nu_{f_L}$ and $\bar N_{f_R}~(f_R=A,B)$.
Although the state identification is still not done at this stage, 
hierarchies among the above mass parameters are assumed as\footnote{
The inequality between $m_1$ and $m_2$ is reversed from the one in
Ref. \cite{sue1}. The present one should be used to make the MSW
mechanism applicable in the $(\psi_1, \psi_2)$ when the state
identification is assumed such as $\psi_1\equiv \bar\nu_{s_1}$ and
$\psi_2\equiv\nu_e$ which will be used in this paper. 
Some related equations in
Ref. \cite{sue1} should be replaced by the ones presented in this paper.}
\begin{equation}
m_{ff^\prime} \ll m_{1} \ll m_{2} \ll m_{3} \sim m_{4} \ll m_{5} \ll M. 
\end{equation}
Since $m_{ff^\prime}$ is small enough compared with others,
we can neglect it for a while.
As a result of the seesaw mechanism, a heavy right-handed neutrino $N_C$
decouples from other neutrinos and a mass matrix for five light states 
$\psi_f$
becomes\footnote{ It should be noted that we approximately call these
states as $\psi_f$. They are not pure $\psi_f$'s but have
a small mixture component from $N_C$.} 
\cite{sue1}
\begin{equation}
{\cal M}_0=M\left( \begin{array}{ccccc}
\mu_1^2 & \mu_1\mu_2 & \mu_1\mu_3 & \mu_1\mu_4 & \mu_1\mu_5 \\
\mu_1\mu_2 & \mu_2^2 & \mu_2\mu_3 & \mu_2\mu_4 & \mu_2\mu_5 \\
\mu_1\mu_3 & \mu_2\mu_3 & \mu_3^2 & \mu_3\mu_4 & \mu_3\mu_5 \\
\mu_1\mu_4 & \mu_2\mu_4 & \mu_3\mu_4 & \mu_4^2 & \mu_4\mu_5 \\
\mu_1\mu_5 & \mu_2\mu_5 & \mu_3\mu_5 & \mu_4\mu_5 & \mu_5^2 \\
\end{array}
\right),
\end{equation} 
where $\mu_f= m_f/M$. As is easily checked, ${\cal M}_0$ is 
a matrix with a rank one and
diagonalized as $U^{(\nu)}{\cal M}_0U^{(\nu)T}$ by using the matrix
\begin{equation}
U^{(\nu)}=\left(\begin{array}{cc}
{\cal O} & 0\\  0 & 1\\ \end{array}\right)
\left(
\begin{array}{ccccc}
\displaystyle{{\mu_2 \over \xi_1}} & \displaystyle{-{\mu_1 \over
\xi_1}} & 0 & 0 & 0 \\
\displaystyle{{\mu_1\mu_3 \over \xi_1\xi_2}} & 
\displaystyle{{\mu_2\mu_3 \over \xi_1\xi_2}} &
\displaystyle{-{\xi_1 \over \xi_2}} & 0 & 0 \\
\displaystyle{{\mu_1\mu_4 \over \xi_2\xi_3}} & 
\displaystyle{{\mu_2\mu_4 \over \xi_2\xi_3}} &
\displaystyle{{\mu_3\mu_4 \over \xi_2\xi_3}} & 
\displaystyle{-{\xi_2 \over \xi_3}} & 0 \\
\displaystyle{{\mu_1\mu_5 \over \xi_3\xi_4}} & 
\displaystyle{{\mu_2\mu_5 \over \xi_3\xi_4}} &
\displaystyle{{\mu_3\mu_5 \over \xi_3\xi_4}} & 
\displaystyle{{\mu_4\mu_5 \over \xi_3\xi_4}} & 
\displaystyle{-{\xi_3 \over \xi_4}} \\
\displaystyle{{\mu_1 \over \xi_4}} & 
\displaystyle{{\mu_2 \over \xi_4}} & 
\displaystyle{{\mu_3 \over \xi_4}} &
\displaystyle{{\mu_4 \over \xi_4}} & 
\displaystyle{{\mu_5 \over \xi_4}}
\end{array} \right),
\end{equation}
where $\displaystyle \xi_n^2=\sum_{f=1}^{n+1}\mu_f^2$.
${\cal O}$ is an undetermined $4\times 4$ matrix which should be 
introduced because of the mass degeneracy.

To resolve this mass degeneracy and fix $U^{(\nu)}$ completely, 
we now switch on rather complicated mass corrections $m_{ff^\prime}$ 
to the five light
neutrinos in order to yield the hierarchical mass eigenvalues
without disturbing the mixing structure $U^{(\nu)}$.
As such masses we consider the simplest example as 
\begin{equation}
{\cal M}_{\rm per}
\simeq
\left( \begin{array}{ccccc}
{\cal A}\mu_1^2 & {\cal B}\mu_1\mu_2 & {\cal D}\mu_1\mu_3 & 
{\cal E}\mu_1\mu_4 & {\cal F}\mu_1\mu_5\\
{\cal B}\mu_1\mu_2 & {\cal C}\mu_2^2 & {\cal D}\mu_2\mu_3 & 
{\cal E}\mu_2\mu_4 & {\cal F}\mu_2\mu_5\\
{\cal D}\mu_1\mu_3 & {\cal D}\mu_2\mu_3 & {\cal D}\mu_3^2 & 
{\cal E}\mu_3\mu_4 & {\cal F}\mu_3\mu_5\\
{\cal E}\mu_1\mu_4 & {\cal E}\mu_2\mu_4 & {\cal E}\mu_3\mu_4 & 
{\cal D}\mu_4^2 & {\cal F}\mu_4\mu_5 \\
{\cal F}\mu_1\mu_5 & {\cal F}\mu_2\mu_5 & {\cal F}\mu_3\mu_5 &
{\cal F}\mu_4\mu_5 & {\cal G}\mu_5^2 \\
\end{array}
\right),
\end{equation}
where ${\cal A} \sim {\cal F}$ are parameters which satisfy
\begin{equation}
{{\cal A}-{\cal D}\over {\cal B}-{\cal D}}=
{{\cal B}-{\cal D}\over {\cal C}-{\cal D}}=-{\mu_2^2\over\mu_1^2},\qquad
{{\cal F}\over {\cal D}+{\cal E}}=-{\mu_4^2 \over \mu_5^2}.
\end{equation}
This matrix can also be checked to be diagonalized by $U^{(\nu)}$ and the 
mass eigenvalues of 
${\cal M}_0+ {\cal M}_{\rm per}$ is obtained as
\begin{equation}
M_{\bar 1}=({\cal A}-{\cal D})\mu_1^2,~ M_{\bar 2}=0, ~
M_{\bar 3}=({\cal D}-{\cal E})\mu_3^2,~
M_{\bar 4}=({\cal D}+{\cal E})\mu_4^2,~M_{\bar 5}=M\mu_5^2.
\end{equation}
Hereafter we use numerical indicies with a bar to specify the mass
eigenstates.  
A 55-element of ${\cal M}_{\rm per}$ can take its value in a range like
$\vert {\cal G}\vert~{^<_\sim}~O( \vert {\cal F}\vert) $ to guarantee the 
approximate diagonalization by $U^{(\nu)}$ because the fifth mass eigenvalue 
$M_{\bar 5}$ coming from ${\cal M}_0$ is rather large.
After the introduction of this correction the mass degeneracy is 
completely resolved and ${\cal O}$ in Eq. (11)
is determined as ${\cal O}={\bf 1}$.

Here it is useful to note how many number of parameters are included 
in this model.
As easily found from Eqs. (11) and (14), there are nine free
parameters. They may be taken as $\mu_f (f=1\sim 5), 
M, {\cal C}, {\cal D}$ and ${\cal E}$.
It is convenient for the later study to present concrete expressions of 
oscillation parameters in Eq. (7), in particular, 
the amplitude $-4V_{\alpha f^\prime}V_{\alpha f}
V_{\beta f^\prime}V_{\beta f}$ by using these parameters\footnote{ 
For a while, we confine our attention to $U^{(\nu)}$ 
assuming that a charged lepton mass matrix is diagonal so that
$U^{(l)}={\bf 1}$.
In section 4.3 we extend our analysis to more general cases.}.
For the $\psi_f\leftrightarrow\psi_{f^\prime}$ oscillation their 
expressions due to the $\alpha\beta$-sector are presented in Table 1.
Although there are small negative contributions to ${\cal
P}_{\psi_{f^\prime}\rightarrow\psi_f}$ from some $\alpha\beta$-sectors, we omit
them from Table 1. 
The detailed explanation of numerical values listed in the columns (I) and
(II) of Table 1 is given in the next section.

\begin{figure}[hbt]
\begin{center}
\begin{tabular}{c|ccc|ccc}
$(\psi_f,\psi_{f^\prime})$ &\multicolumn{3}{c|}{$\Delta M^2_{\alpha\beta}$} &
\multicolumn{3}{c}{$-4V_{\alpha f^\prime}V_{\alpha f}
V_{\beta f^\prime}V_{\beta f}$}
\\ \cline{2-4}\cline{5-7}
 & & (I) & (II) & & \multicolumn{2}{c}{(I), (II)} \\
\hline\hline
$(\nu_{s_1},\nu_e)$& $\Delta M^2_{\bar 1\bar 2}$ & $10^{-5}$ & $10^{-5}$ &
$4({\mu_1\over \mu_2})^2$ & \multicolumn{2}{c}{$10^{-2.2}$}\\  
 & $\Delta M^2_{\bar 1\bar 3}$ & $10^{-5}$ & 1 &
$2({\mu_1\over \mu_3})^2$ & \multicolumn{2}{c}{$10^{-2a-2.5}$}\\  
 & $\Delta M^2_{\bar 1\bar 4}$ & $10^{-2.4}$ & 1 &
$2({\mu_1\over \mu_3})^2$ & \multicolumn{2}{c}{$10^{-2a-2.5}$}\\  
 & $\Delta M^2_{\bar 1\bar 5}$ & $M_{\bar 5}^2$ & $M_{\bar 5}^2$ &
$4({\mu_1\over \mu_5})^2$ & \multicolumn{2}{c}{$10^{-2(a+b)-2.2}$}\\ \hline
$(\nu_{s_1},\nu_\mu)$& $\Delta M^2_{\bar 2\bar 3}$ & $10^{-5}$ &1 &
 $2({\mu_1\over \mu_3})^2$ & 
\multicolumn{2}{c}{$10^{-2a-2.5}$}\\
 & $\Delta M^2_{\bar 2\bar 4}$ & $10^{-2.4}$ &1 &
 $2({\mu_1\over \mu_3})^2$ & \multicolumn{2}{c}{$10^{-2a-2.5}$}\\
 & $\Delta M^2_{\bar 2\bar 5}$ &$M_{\bar 5}^2$  & $M_{\bar 5}^2$ &
 $4({\mu_1\over \mu_5})^2$ & \multicolumn{2}{c}{$10^{-2(a+b)-2.2}$}\\ \hline
$(\nu_{s_1},\nu_\tau)$& $\Delta M^2_{\bar 3\bar 4}$ & $10^{-2.4}$ & 
$10^{-2.4}$ & $({\mu_1\over\mu_3})^2$
&\multicolumn{2}{c}{$10^{-2a-2.8}$} \\
 & $\Delta M^2_{\bar 3\bar 5}$ & $M_{\bar 5}^2$ & $M_{\bar 5}^2-1$
& $2({\mu_1\over\mu_5})^2$ &\multicolumn{2}{c}{$10^{-2(a+b)-2.5}$} \\ \hline
$(\nu_{s_1},\nu_{s_2})$& $\Delta M^2_{\bar 4\bar 5}$ & $M_{\bar 5}^2$ 
& $M_{\bar 5}^2-1$ & $4({\mu_1\over\mu_5})^2$ 
&\multicolumn{2}{c}{$10^{-2(a+b)-2.2}$} \\ \hline
$(\nu_e,\nu_\mu)$& $\Delta M^2_{\bar 2\bar 3}$ & $<10^{-2.4}$ &$ 1$& 
$2({\mu_2\over\mu_3})^2$ &\multicolumn{2}{c}{$10^{-2a+0.3}$}\\
  & $\Delta M^2_{\bar 2\bar 4}$ &$10^{-2.4}$  &$ 1$& 
$2({\mu_2\over\mu_3})^2$ &\multicolumn{2}{c}{$10^{-2a+0.3}$}\\
     & $\Delta M^2_{\bar 2\bar 5}$ &$M_{\bar 5}^2$  & $M_{\bar 5}^2$ &
$4({\mu_2\over\mu_5})^2$ &\multicolumn{2}{c}{$10^{-2(a+b)+0.6}$} \\\hline
$(\nu_e,\nu_\tau)$& $\Delta M^2_{\bar 3\bar 4}$ & 
$10^{-2.4}$ &$10^{-2.4}$  & $ ({\mu_2\over\mu_3})^2$ 
&\multicolumn{2}{c}{$10^{-2a}$} \\
     & $\Delta M^2_{\bar 3\bar 5}$ & $M_{\bar 5}^2$ & $M_{\bar 5}^2-1$&
     $2({\mu_2\over\mu_5})^2$ &
\multicolumn{2}{c}{$10^{-2(a+b)+0.3}$} \\ \hline
$(\nu_e,\nu_{s_2})$& $\Delta M^2_{\bar 4\bar 5}$ & $M_{\bar 5}^2$ 
&$M_{\bar 5}^2-1$ & $4({\mu_2\over\mu_5})^2$ 
&\multicolumn{2}{c}{$10^{-2(a+b)+0.6}$} \\ \hline
$(\nu_\mu,\nu_\tau)$& $\Delta M^2_{\bar 3\bar 4}$ & $10^{-2.4}$ &$10^{-2.4}$& 
$4({\mu_3\over\mu_4})^2(1+({\mu_3\over\mu_4})^2)^{-2}$   &
\multicolumn{2}{c}{1} \\
&$\Delta M^2_{\bar 3\bar 5}$ &$M_{\bar 5}^2$  &$M_{\bar 5}^2-1$& 
$2({\mu_3\over\mu_5})^2$  &
\multicolumn{2}{c}{$10^{-2b+0.3}$} \\ \hline
$(\nu_\mu,\nu_{s_2})$& $\Delta M^2_{\bar 4\bar 5}$ &$M_{\bar 5}^2$ 
&$M_{\bar 5}^2-1$& $4({\mu_3\over\mu_5})^2$  &
\multicolumn{2}{c}{$10^{-2b+0.6}$} \\ \hline
$(\nu_\tau,\nu_{s_2})$& $\Delta M^2_{\bar 4\bar 5}$ & $M_{\bar 5}^2$ &
$M_{\bar 5}^2-1$& 
$4({\mu_3\over\mu_5})^2$  &
\multicolumn{2}{c}{$10^{-2b+0.6}$} \\ \hline
\end{tabular}\vspace{4mm}\\
\end{center}
{\bf Table 1}~~ Oscillation parameters for $\psi_f \leftrightarrow
\psi_{f^\prime}$. The positive contribution to the oscillation probability
 does not come from other combinations of $\alpha$ and $\beta$. 
In this table we assume $U^{(l)}={\bf 1}$.  
The state identification $\psi_f=(\bar N_A(\equiv
s_1),~\nu_{f_L},~\bar N_B(\equiv s_2))$ is assumed.
\end{figure}

\subsection{Construction based on nonrenormalizable terms}
In the previous subsection we introduced our mass matrix ${\cal
M}_0+{\cal M}_{\rm per}$ in the phenomenological way 
so as to realize the substantial mixing between sterile 
and active neutrinos with nondegenerating masses. 
However, we can also construct it under a certain theoretical
background. We present such an example in this subsection.
To make our argument definite we consider a certain type of
supersymmetric nonanomalous extra U(1) models with a 
D-flat direction for the extra U(1), which comes from $E_6$
unification group \cite{esix}.
This kind of models may be expected to be induced from 
the perturbative superstring. 
The following argument will be straightforwardly applied to other type of
models, for example, anomalous U(1) models.

This example contains three right-handed neutrinos $\bar N_{f_R}$
and pairs of singlets $(S_\ell,\bar S_\ell)$ $(\ell=1,2)$, where 
$\bar N_{f_R}$ and
$S_\ell$ belong to {\bf 27} chiral superfield of $E_6$ and a conjugate 
partner $\bar S_\ell$ belongs to ${\bf 27}^*$ chiral superfield. 
The extension to the case with $\ell \ge 3$ is straightforward.
$\bar N_{f_R}$, $S_\ell$ and $\bar S_\ell$ are singlets
under the gauge groups of the standard model but have the extra U(1) charge.
$\bar N_{f_R}$ and $S_\ell$ are assumed to have the same charge of
this extra U(1).
The gauge invariant relevant terms in the superpotential to the 
present investigation are
composed by two parts,
\begin{eqnarray}
W_0&=& \sum_{\ell=1,2}{k_{K_\ell}\over M_G^{2K_\ell-3}} 
(S_\ell\bar S_\ell)^{K_\ell}, \\
W_1&=&\sum_{f_R,f_R^\prime}
{k_S^{f_Rf_R^\prime} \over M_G^{2P_{f_Rf_R^\prime}-3}}
\left(S_2\bar S_2\right)^{p_{f_Rf_R^\prime}} 
\left(S_1\bar S_1\right)^{P_{f_Rf_R^\prime}-p_{f_Rf_R^\prime}-2} 
\bar S_1^2 \bar N_{f_R} \bar N_{f_R^\prime}  \nonumber \\
\hspace*{1cm}&&+ \sum_{f_L,f_R}{k_D^{f_Lf_R}\over M_G^{2Q_{f_Lf_R}}}
\left( S_2\bar S_2\right)^{q_{f_Lf_R}} 
(S_1\bar S_1)^{Q_{f_Lf_R}-q_{f_Lf_R}}L_{f_L}H_2\bar N_{f_R} \nonumber \\ 
\hspace*{1cm}&&+ \sum_{f_L,f_L^\prime} 
{k_M^{f_Lf_L^\prime}\over M_G^{2R_{f_Lf_L^\prime}+3}} 
\left( S_2\bar S_2\right)^{r_{f_Lf_L\prime}}
\left(S_1\bar S_1\right)^{R_{f_Lf_L^\prime}-r_{f_Lf_L^\prime}}
S_1^2 H_2^2 L_{f_L}L_{f_L^\prime},
\end{eqnarray}
where the parameters determining the power structure of each term are
integers and satisfy 
$$K_\ell \ge 2,\quad P_{f_Rf^\prime_R}-2\ge p_{f_Rf^\prime_R} \ge 0,\quad
Q_{f_Lf_R} \ge q_{f_Lf_R} \ge 0,\quad 
R_{f_Lf_L^\prime}\ge r_{f_Lf_L^\prime}\ge 0.$$
$M_G$ is a suitable scale such as a string scale or a Planck scale. 
The higher order terms can be neglected in comparison with these
lowest dimensional terms.
The scalar potential for the singlet scalars $S_\ell$ and $\bar S_\ell$ 
can be written as
\begin{eqnarray}
V&=&{g_x^2 \over 2}\sum_{\ell=1,2} Q_{S_\ell}^2\left(
\vert S_\ell\vert^2- \vert \bar S_\ell\vert^2 \right)^2 \nonumber \\
&+& \sum_{\ell=1,2}\left\{\vert {k_{K_\ell} K_\ell S_\ell^{K_\ell-1}\bar
S_\ell^{K_\ell} 
\over M_G^{2K_\ell-3}}\vert^2
+ \vert {k_{K_\ell} K_\ell S_\ell^{K_\ell}\bar S_\ell^{K_\ell-1} \over 
M_G^{2K_\ell-3}}\vert^2
- m^2\left( \vert S_\ell\vert^2 
+ \vert\bar S_\ell\vert^2\right)\right\}.
\end{eqnarray}
The first line is a D-term contribution of the extra U(1) 
and the second line represents
an F-term contribution coming from $W_0$ and  soft 
supersymmetry breaking scalar masses which we assumed as 
$m^2_{S_\ell}=m^2_{\bar S_\ell}= m^2$, for simplicity.
Clearly this scalar potential has D-flat directions 
$\vert S_\ell\vert=\vert\bar S_\ell\vert\equiv u_\ell$ along which 
the potential minimum is realized.
After soft supersymmetry breaking masses are introduced as in Eq. (17), 
the magnitude of the intermidiate scale $u_\ell$ is determined by the
nonrenormalizable F-term contribution in the second line as
\begin{equation}
u_\ell=\left({1 \over (2K_\ell-1)^{1/2}k_{K_\ell} K_\ell}
mM_G^{2K_\ell-3}\right)^{1\over 2K_\ell-2},
\end{equation}
where the value of $u_\ell$ is crucially dependent on an integer $K_\ell$.
Moreover, as easily seen in Eq. (16), depending on the values of 
powers in the nonrenormalizable terms such as 
$P_{f_Rf_R^\prime}$, $Q_{f_Lf_R^\prime}$ and 
$R_{f_Lf_L^\prime}$, 
various types of neutrino mass terms 
can be induced through the effective couplings
controled by the power of $u_\ell/M_G$ 
after the vacuum expectation value $u_\ell$ becomes nonzero.

Since there is no restriction on the right-handed neutrino mass 
terms due to the gauge symmetry unlike the usual Grand Unified models, 
this framework can 
contain light sterile neutrinos in addtion to 
the ordinary three active light neutrinos.
If we take the basis $\psi_f$ in Eq. (8) as $\psi_f=(\bar
N_A(\equiv\nu_{s_1}),~ \nu_{f_L},~\bar
N_B(\equiv\nu_{s_2}))$,\footnote{The phenomenological 
validity of this identification
will be discussed in the next section.} 
the mass parameters in Eq. (8) can be written by using the parameters
in Eq. (16) as 
\begin{eqnarray}
&&m_1=k_S^{AC}\epsilon^{2P_{AC}-3}\delta^{2p_{AC}}u_1 ,\quad
m_{f_L}=k_D^{f_L C}\epsilon^{2Q_{f_L C}}
\delta^{2q_{f_L C}}v_2, \nonumber \\
&&m_5=k_S^{BC}\epsilon^{2P_{BC}-3}\delta^{2p_{BC}}u_1, \quad
M=k_S^{CC}\epsilon^{2P_{CC}-3}\delta^{2p_{CC}}u_1,
\end{eqnarray}
where we use the definitions $\epsilon\equiv u_1/M_G$ and
$\delta\equiv u_2/u_1$. In Eq. (19) the indicies
$f_L=e, \mu, \tau$ should be interpreted to correspond to the numerical
indices 2, 3 and 4 in Eq. (8).
Then the parameters $\mu_f$ are also expressed as
\begin{eqnarray}
&&\mu_1={k_S^{AC} \over k_S^{CC}}
\epsilon^{2(P_{AC}-P_{CC})}\delta^{2(p_{AC}-p_{CC})}, \quad
\mu_{f_L}={k_D^{f_LC} \over k_S^{CC}}{v_2\over u_1}
\epsilon^{2(Q_{f_LC}-P_{CC})+3}\delta^{2(q_{f_L C}-p_{CC})}, \nonumber \\
&&\mu_5={k_S^{BC} \over k_S^{CC}}
\epsilon^{2(P_{BC}-P_{CC})}\delta^{2(p_{BC}-p_{CC})}.
\end{eqnarray}
Moreover, based on the superpotential $W_1$ 
we can write the neutrino mass matrix ${\cal M}_{\rm per}$ 
under the same basis as follows,
\begin{equation}
{\cal M}_{\rm per}=\left(
\begin{array}{lcl}
k_S^{AA}\epsilon^{2P_{AA}-3}\delta^{2p_{AA}}u_1&
k_D^{f_LA}\epsilon^{2Q_{f_L A}}\delta^{2q_{f_L A}}v_2 &
k_S^{AB}\epsilon^{2P_{AB}-3}\delta^{2p_{AB}}u_1 \\
k_D^{f_LA}\epsilon^{2Q_{f_L A}}\delta^{2q_{f_L A}}v_2 &
{\cal M}_{f_Lf_L^\prime} &
k_D^{f_LB}\epsilon^{2Q_{f_L B}}\delta^{2q_{f_L B}}v_2\\
k_S^{AB}\epsilon^{2P_{AB}-3}\delta^{2p_{AB}}u_1 &
k_D^{f_LB}\epsilon^{2Q_{f_L B}}\delta^{2q_{f_L B}}v_2 &
k_S^{BB}\epsilon^{2P_{BB}-3}\delta^{2p_{BB}}u_1
\end{array}\right),
\end{equation}
where $v_2$ is a VEV of the doublet Higgs scalar $H_2$.
Majorana masses ${\cal M}_{f_Lf_L^\prime}$ can be caused by the last
terms in Eq. (16). However, even if $R_{f_Lf_L^\prime} =0$, 
these Majorana masses are $\sim 10^{-7}$ eV and then generally 
too small for the explanation of the solar and atmospheric 
neutrino problem \cite{lang}.
Within the interactions included in Eq. (16) 
${\cal M}_{f_Lf_L^\prime}$ can be also induced
as the mixing terms composed by the first and the second terms of
Eq. (16). But it is difficult to arrange in the way 
to satisfy the condition given by Eq. (13) 
and guarantee the sufficient largeness of eigenvalues.
Finally we are forced to introduce new interaction terms.
Here we introduce a triplet Higgs scalar $\Phi$ and consider the following
interaction terms in the superpotential \footnote{
If we try to construct the model including a triplet Higgs scalar in
the perturbative superstring, it must be necessary to consider the
models with higher Kac-Moody level.}: 
\begin{equation}
W_2={k_T^{f_Lf_L^\prime}\over M_G^{2T_{f_Lf_L^\prime}}}
(S_2\bar S_2)^{t_{f_Lf_L^\prime}}
(S_1\bar S_1)^{T_{f_Lf_L^\prime}-t_{f_Lf_L^\prime}}\Phi L_{f_L}L_{f_L^\prime}.
\end{equation}
This gives the Majorana masses 
\begin{equation}
{\cal M}_{f_Lf_L^\prime}=
k_T^{f_Lf_L^\prime}\epsilon^{2T_{f_Lf_L^\prime}}
\delta^{t_{f_Lf_L^\prime}}v_T,
\end{equation}
where $v_T$ is a VEV of $\Phi$. There is a constraint on $v_T$ from
an electroweak $\rho$ parameter. To satisfy its constraint
it will be necessary to take it as $v_T~{^<_\sim}~ 1$~GeV. 

What kind of the mass matrix ${\cal M}_{\rm per}$ is induced in this model 
is completely dependent on the values of the lowest powers and couplings of 
each terms in the superpotential shown by Eqs. (15), (16) and
(22).
In order to constrain the superpotential, 
it is necessary to be able to introduce a certain kind of symmetry
which can forbid the lower dimensional terms consistently. 
In many works \cite{matrix} an Abelian horizontal symmetry has been  
used to constrain the nonrenormalizable superpotential, which can 
induce the small neutrino masses with a favorable texture after the
breakdown of this horizontal symmetry due to the VEV of some scalar fields. 
In our model the discrete symmetry may play the same role.
To realize the superpotential starting from high dimensional terms, 
we need rather complicated higher order discrete symmetry.
Here we assume $Z_9\times Z_9$ as a concrete example of such 
a discrete symmetry and take the charge assignments for the relevant 
fields under this symmetry as the ones shown in Table 2.
The charge assignment for this discrete symmetry allows the lowest terms
with $K_1=K_2=9$ in $W_0$. 
This results in $u_1\sim u_2\sim 10^{17}$~GeV and then 
$\epsilon\sim 0.1$ and $\delta\sim 1$.
We should note here that in this case $\epsilon$ is not so 
small that the extremely high dimensional terms are necessary to 
realize the desirable neutrino masses.
If we adopt a smaller $K_\ell$ and make $\epsilon$ small enough, we
do not need such high dimensional terms but additional fine tunings
for coupling coefficients are required to induce a necessary mass
pattern. 
In such a case the main feature of the mass pattern is determined 
by the tunings of the couplings. Thus we do not take this way here.

Terms in $W_1$ and $W_2$ are also constrained by this discrete
symmetry and the following values for the powers in $W_1$ and $W_2$
are allowed as the lowest ones:
\begin{eqnarray}
&&P_{f_Rf_R^\prime}=\left(
\begin{array}{ccc}
18&17&11\\ 17&16&10\\ 11&10&4
\end{array}\right), \qquad
Q_{f_Lf_R}=\left(
\begin{array}{ccc}
9 & 8& 2\\ 8&7&1\\ 8&7&1
\end{array} \right), \qquad
T_{f_Lf_L^\prime}=\left(
\begin{array}{ccc}
8 &7 &7 \\7&6& 6\\ 7&6&6
\end{array} \right),
\nonumber \\
&&p_{f_Rf_R^\prime}=\left(
\begin{array}{ccc}
8&8&4\\ 8&8&4\\ 4&4&0
\end{array}\right), \qquad
q_{f_Lf_R}=\left(
\begin{array}{ccc}
4 &4 &0 \\4&4& 0\\ 4&4&0
\end{array} \right),\qquad
t_{f_Lf_L^\prime}=\left(
\begin{array}{ccc}
6 &6 &6 \\6&6& 6\\ 6&6&6
\end{array} \right).
\end{eqnarray}
The present examples may not be considered to be realistic
because of their extremely large dimensions of the necessary
nonrenormalizable terms. Although this feature seems to be general in
this type of models, 
the situation may be made mild to some extent
by considering a type of extra U(1) models in which a role of 
$S_\ell\bar S_\ell$ is
replaced by suitable elementary singlet fields with D-flat directions.
Anyway this example shows the direction how we can constrain the
nonrenormalizable terms in the superpotential.

Next we study the neutrino mass matrix induced by this superpotential. 
If we take $v_2\sim\epsilon^{15}u_1$, $v_T\sim 1$~GeV, these values give 
$\mu_f$ and ${\cal M}_{\rm per}$ as follows:
\begin{eqnarray}
&&\mu_1= \tilde k_S^{AC}\epsilon^{14}, \quad 
\mu_2= \tilde k_D^{eC}\epsilon^{14}, \quad 
\mu_3= \tilde k_D^{\mu C}\epsilon^{12}, \quad
\mu_4= \tilde k_D^{\tau C}\epsilon^{12}, \quad 
\mu_5= \tilde k_S^{BC}\epsilon^{12}, \nonumber\\
&& \nonumber\\
&&
\left( \begin{array}{lllll}
{\displaystyle {\tilde k_S^{AA}\over (\tilde k_S^{AC})^2}\mu_1^2} &
{\displaystyle {\tilde k_D^{eA}\over \tilde k_D^{eC}\tilde
k_S^{AC}}\mu_1\mu_2} &
{\displaystyle {\tilde k_D^{\mu A}\over \tilde 
k_D^{\mu C}\tilde k_S^{AC}}\mu_1\mu_3} &
{\displaystyle {\tilde k_D^{\tau A}\over \tilde k_D^{\tau C}\tilde
k_S^{AC}}\mu_1\mu_4} &
{\displaystyle {\tilde k_S^{AB}\over \tilde k_S^{BC}\tilde
k_S^{AC}}\mu_1\mu_5}  \\
{\displaystyle {\tilde k_D^{eA}\over \tilde k_D^{eC}
\tilde k_S^{AC}}\mu_1\mu_2} &
{\displaystyle{\tilde k_T^{ee}\over (k_D^{eC})^2}\mu_2^2} &
{\displaystyle{\tilde k_T^{e\mu}\over \tilde k_D^{e C}\tilde k_D^{\mu
C}}\mu_2\mu_3}  &
{\displaystyle{\tilde k_T^{e\tau}\over \tilde k_D^{e C}\tilde
k_D^{\tau C}}\mu_2\mu_4}  &
{\displaystyle{\tilde k_D^{eB}\over \tilde k_D^{eC}\tilde
k_S^{BC}}\mu_2\mu_5}  \\
{\displaystyle{\tilde k_D^{\mu A}\over \tilde k_D^{\mu C}
\tilde k_S^{AC}}\mu_1\mu_3}  &
{\displaystyle{\tilde k_T^{e\mu}\over \tilde k_D^{e C}\tilde k_D^{\mu
C}}\mu_2\mu_3}  &
{\displaystyle{\tilde k_T^{\mu\mu}\over (k_D^{\mu C})^2}\mu_3^2} &
{\displaystyle{\tilde k_T^{\mu\tau}\over \tilde k_D^{\mu C}\tilde
k_D^{\tau C}} \mu_3\mu_4} &
{\displaystyle{\tilde k_D^{\mu B}\over \tilde k_D^{\mu C}\tilde
k_S^{BC}}\mu_3\mu_5}  \\
{\displaystyle{\tilde k_D^{\tau A}\over \tilde k_D^{\tau C}\tilde
k_S^{AC}}\mu_1\mu_4} &
{\displaystyle{\tilde k_T^{e\tau}\over \tilde k_D^{e C}\tilde
k_D^{\tau C}}\mu_2\mu_4}  &
{\displaystyle{\tilde k_T^{\mu\tau}\over \tilde k_D^{\mu C}\tilde
k_D^{\tau C}}\mu_3\mu_4}  &
{\displaystyle{\tilde k_T^{\tau\tau}\over (k_D^{\tau C})^2}\mu_4^2} &
{\displaystyle{\tilde k_D^{\tau B}\over \tilde k_D^{\tau C}\tilde
k_S^{BC}}\mu_4\mu_5}  \\
{\displaystyle{\tilde k_S^{AB}\over \tilde k_S^{BC}\tilde
k_S^{AC}}\mu_1\mu_5}  &
{\displaystyle{\tilde k_D^{eB}\over \tilde k_D^{eC}\tilde
k_S^{BC}}\mu_2\mu_5}  &
{\displaystyle{\tilde k_D^{\mu B}\over \tilde k_D^{\mu C}\tilde
k_S^{BC}}\mu_3\mu_5}  &
{\displaystyle{\tilde k_D^{\tau B}\over \tilde k_D^{\tau C}\tilde
k_S^{BC}}\mu_4\mu_5 } &
{\displaystyle{\tilde k_S^{BB}\over (\tilde k_S^{BC})^2}\mu_5^2 } \\
\end{array}\right), \nonumber \\
\end{eqnarray}
where $\tilde k_x^{ff^\prime}\equiv k_x^{ff^\prime}/ k_S^{CC}$.
To obtain the phenomenologically favorable hierarchy 
among $\mu_f ~(f=1\sim 5)$ 
and realize the mass matrix form defined by Eqs. (12) and (13), 
we need additional hierarchical structure in the coupling constants 
$\tilde k_x^{ff^\prime}$.
Among these tunings of coupling constants 
we introduce the following useful parameters expressing the freedom
which can not be determined by the present experimental results in our 
state identification used later:
\begin{equation}
a=2-\log_{10}\left({k_D^{e C}\over k_D^{\mu C}}\right), \qquad
b=-\log_{10}\left({k_D^{\tau C}\over k_S^{BC}}\right).
\end{equation} 
Using these parameters and if we make the suitable assumptions on 
$\tilde k_x^{ff^\prime}$, $\mu_f$ can have the following hierarchy:
\begin{equation}
\mu_1\sim \epsilon^{16}, \quad 
\mu_2\sim \epsilon^{12+a}, \quad 
\mu_3\sim \epsilon^{12}, \quad
\mu_4\sim \epsilon^{10+b}, \quad 
\mu_5\sim \epsilon^{10}. 
\end{equation}
Although there are many possibilities for ${\cal M}_{\rm per}$ 
depending on the assumption on the coupling constants,
it will be useful to present examples to see what kind of tunings
of the coupling constants are required to construct the interesting
${\cal M}_{\rm per}$. Here we give two examples: 
\begin{eqnarray}
{\rm (I)}: && {\cal M}_{\rm per}\sim M \left( \begin{array}{ccccc}
\epsilon^{-6.3}\mu_1^2 & 
-\epsilon^{-3.5}\mu_1\mu_2 & \epsilon^{-1.9}\mu_1\mu_3 & \epsilon^{-1.9}
\mu_1\mu_4 & -\epsilon^{2.2}\mu_1\mu_5\\
-\epsilon^{-3.5}\mu_1\mu_2 & \epsilon^{-1.9}\mu_2^2 & \epsilon^{-1.9}
\mu_2\mu_3 & \epsilon^{-1.9}\mu_2\mu_4 &-\epsilon^{2.2}\mu_2\mu_5\\
\epsilon^{-1.9}\mu_1\mu_3 & \epsilon^{-1.9}\mu_2\mu_3 &
\epsilon^{-1.9}\mu_3^2 & \epsilon^{-1.9}\mu_3\mu_4 & 
-\epsilon^{2.2}\mu_3\mu_5\\
\epsilon^{-1.9}\mu_1\mu_4 & \epsilon^{-1.9}\mu_2\mu_4 & \epsilon^{-1.9}
\mu_3\mu_4 & \epsilon^{-1.9}\mu_4^2 & -\epsilon^{2.2}\mu_4\mu_5 \\
-\epsilon^{2.2}\mu_1\mu_5 & -\epsilon^{2.2}\mu_2\mu_5 & 
-\epsilon^{2.2}\mu_3\mu_5 &
-\epsilon^{2.2}\mu_4\mu_5 & \epsilon^{2.2}\mu_5^2 \\
\end{array}
\right), \nonumber \\
{\rm (II)}: && {\cal M}_{\rm per}\sim M \left( \begin{array}{ccccc}
\epsilon^{-6.3}\mu_1^2 & 
-\epsilon^{-3.5}\mu_1\mu_2 & \epsilon^{-3.4}\mu_1\mu_3 &
\epsilon^{0.4} \mu_1\mu_4 & 
-\epsilon\mu_1\mu_5\\
-\epsilon^{-3.5}\mu_1\mu_2 & \epsilon^{-3.4}\mu_2^2 & 
\epsilon^{-3.4}\mu_2\mu_3 & 
\epsilon^{0.4}\mu_2\mu_4 & -\epsilon\mu_2\mu_5\\
\epsilon^{-3.4}\mu_1\mu_3 & \epsilon^{-3.4}\mu_2\mu_3 & \epsilon^{-3.4}
\mu_3^2 & \epsilon^{0.4}\mu_3\mu_4 & -\epsilon\mu_3\mu_5\\
\epsilon\mu_1\mu_4 & \epsilon\mu_2\mu_4 & \epsilon\mu_3\mu_4 & 
\epsilon^{-3.4}\mu_4^2 & 
-\epsilon\mu_4\mu_5 \\
-\epsilon\mu_1\mu_5 & -\epsilon\mu_2\mu_5 & -\epsilon\mu_3\mu_5 &
-\epsilon\mu_4\mu_5 & \epsilon\mu_5^2 \\
\end{array}\right).
\end{eqnarray}
These examples have the interesting penomenological 
features as explained in the next section. 
Although we need additional tunings for the coupling constants to
satisfy the condition shown by Eq. (13) as found from these examples, 
such tunings seem not to be so hard but rather mild as we can find by 
comparing Eqs. (25) and (28). We believe that it does not 
spoil the interesting feature of the present model. 

For the charged lepton sector the same discrete symmetry can also determine
the structure of the nonrenormalizable terms in the superpotential
\begin{equation}
W_3=\sum_{f_L,f_R} {k_E^{f_Lf_R}\over M_G^{2X_{f_Lf_R}}}
(S_2\bar S_2)^{x_{f_Lf_R}}(S_1\bar S_1)^{X_{f_Lf_R}-x_{f_Lf_R}}
L_{f_L}H_1\bar E_{f_R}
\end{equation}
following the charge assignment in Table 2.
By the use of this symmetry we can write down the charged lepton mass matrix
$M^{(\ell)}$ as follows:
\begin{equation}
(\bar e_R, \bar\mu_R, \bar\tau_R)\left( \begin{array}{lll}
k_E^{ee}v_1\epsilon^6 & k_E^{e\mu}v_1\epsilon^{4}&
k_E^{e\tau}v_1\epsilon^{4}\\ k_E^{\mu e}v_1\epsilon^{4} &
k_E^{\mu\mu}v_1\epsilon^2
& k_E^{\mu\tau}v_1\epsilon^2\\
k_E^{\tau e}v_1\epsilon^{4} & k_E^{\tau\mu}v_1\epsilon^2& 
k_E^{\tau\tau}v_1\epsilon^2\\
\end{array} \right)
\left(\begin{array}{c} e_L\\ \mu_L\\ \tau_L \\
\end{array}
\right),
\end{equation}
where coupling constants again need to be tuned to realize the
desirable charged lepton
mass eigenvalues.
Here we assume that $k_E^{\mu\mu} < k_E^{\tau\tau}$ 
and also the off-diagonal couplings are smaller than these diagonal couplings.
Then this mass matrix can be diagonalized by the bi-unitary transformation
and results in the eigenvalues
\begin{equation}
m_e\sim k_E^{ee}v_1\epsilon^6, \qquad
m_\mu\sim k_E^{\mu\mu}v_1\epsilon^2, \qquad
m_\tau\sim k_E^{\tau\tau}v_1\epsilon^2.
\end{equation}
The diagonalization matrix $U^{(l)}$ is almost diagonal and can be
written as
\begin{equation}
U^{(l)}\sim \left(
\begin{array}{ccc} 1 & -(k_E^{e\mu}/k_E^{\mu\mu})\epsilon^2 & 
-(k_E^{e\tau}/k_E^{\tau\tau})\epsilon^2 \\
(k_E^{e\mu}/k_E^{\mu\mu})\epsilon^2 & 1 & -k_E^{\mu\tau}/k_E^{\tau\tau} \\
(k_E^{e\tau}/k_E^{\tau\tau})\epsilon^2  & k_E^{\mu\tau}/k_E^{\tau\tau}& 1 \\ 
\end{array} \right).
\end{equation}

\begin{figure}[hbt]
\begin{center}
\begin{tabular}{lcc|lcc}\hline
Fields & $Z_9$ & $Z_9$ & Fields & $Z_9$ &$Z_9$ \\ \hline\hline 
$S_1$ & $(1- \xi_1)/9$ & 0 & $L_e$ & $-(1+\xi_3)/9$ & 0\\
$\bar S_1$ & $\xi_1/9$ & 0& $L_\mu$ & $-\xi_3/9$ & 0\\
$S_2$ & 0  &$(1-\xi_2)/9$ & $L_\tau$& $-\xi_3/9$ & 0 \\
$\bar S_2$ &0 & $\xi_2/9$ & $\bar E_e$ &$-(2+\xi_1)/9$ & 0\\
$\bar N_A$ & $-(4+\xi_1)/9$ & $-4/9$ & $\bar E_\mu$ &$-(1+\xi_1)/9$& 
0\\
$\bar N_B$ &$-(3+\xi_1 )/9$ & $-4/9$&  $\bar E_\tau$ &$-(1+\xi_1)/9$&0\\
$\bar N_C$ & $-(1+\xi_1)/9$&0 &$\Phi$ & $-2\xi_3/9$ & $-6/9$\\
$H_{1,2}$& $(\xi_1+\xi_3)/9$ & 0 & & & \\\hline
\end{tabular}
\end{center}
\vspace*{3mm}
{\bf Table 2}~~Charge assingments under $Z_9\times Z_9$ discrete symmetry 
for lepton and Higgs sectors.
$L_{f_L}$ and $\bar E_{f_L}$ are the SU(2)$_L$ doublet and singlet 
 lepton chiral superfields, respectively. 
$H_{1,2}$ and $\Phi$ are the usual doublet Higgs and triplet Higgs
chiral superfields. 
$\xi_i~(i=1,2,3)$ are the integers which can give the nontrivial
charges to each fields and satisfy $1\le\xi_i \le 8$. 
\end{figure}

A possibility of the mixings among sterile neutrinos and active
neutrinos has already been proposed based on the nonrenormalizable
interactions in the superpotential in the context of 
superstring inspired models in Ref. \cite{lang}.
Our model may be considered as a concrete example of its realization.
Although our scheme may need some tunings at least for the coupling
constants
in the superpotential to make the mass matrix the required form defined by 
Eqs. (12) and (13) in the precise way, it is interesting that only 
the above values of $P_{f_Rf_R^\prime}$, $Q_{f_Lf_R}$ and
$T_{f_Lf_L^\prime}$ can
approximately make it close to the required form. 
If we assume $u_1\not= u_2$, more available freedom may be applicable.
This kind of models may be recognized as one of many candidates 
for the possible 
neutrino mass matrix realized in the promising supersymmetric models 
inspired by perturbative superstring.

\section{Analysis of various oscillations}
We apply our model to the analysis of neutrino oscillations.
Deficiencies of the solar neutrinos \cite{solar}
and the atmospheric neutrinos \cite{atm,neut98} have been suggested to be 
explained by 
$\nu_e\leftrightarrow\nu_x$ and $\nu_\mu\leftrightarrow\nu_y$ 
oscillations, respectively.
Within a two flavor oscillation framework 
the neutrino squared mass differences and mixing angles 
predicted from these observations are 
for the solar neutrino problem\footnote{
There are also large mixing solutions.
However, we do not consider these solutions in this paper.} \cite{solar,mass1},
\begin{equation}
\Delta m_{\nu_x\nu_e}^2\sim (0.3-1.2)\times 
10^{-5}{\rm eV}^2, \qquad
\sin^22\theta\sim (0.4-1.5)\times 10^{-2},
\end{equation}
and for the atmospheric neutrino problem \cite{atm,neut98},
\begin{equation}
\Delta m_{\nu_y\nu_\mu}^2\sim (4-6)\times
10^{-3}{\rm eV}^2, \qquad
\sin^22\theta ~{^>_\sim}~ 0.85.
\end{equation}
We also take account here that the existence of 
one neutrino species with mass such as $1\sim 10^2$ eV 
has an interesting relevance to the astrophysical observations 
for the large scale structure of the universe. 

In order to present the consistent explanation for
the neutrino mixings shown in Eqs. (33) and (34), we need to identify the 
five light states $\psi_f$ with physical neutrino states.
In this consideration the constraint from the standard big bang
nucleosynthesis (BBN) \cite{bbn} may be useful because it can constrain a
sterile neutrino sector.
The BBN predicts that the effective neutrino species 
during the primordial nucleosynthesis should be less than 3.3.
This fact severely constrains the mixing angle $\theta$ and the squared mass 
difference $\Delta m^2$ between a sterile neutrino $(\nu_s)$ 
and left-handed active neutrinos which mix with it \cite{nucl1,ssf}.
As long as we do not assume the large lepton number asymmetry at the BBN 
epoch \cite{asym2}, these constraints rule out the large mixing MSW 
solution of the solar neutrino
problem due to $\nu_e\rightarrow\nu_{s}$ and also the explanation of the
atmospheric neutrino problem by $\nu_\mu\rightarrow\nu_{s}$.
Taking account of these facts, we concentrate our study on 
a possibility such that 
$\psi_1$ and $ \psi_5$ are right-handed sterile neutrinos $\nu_{s_1}$
and $\nu_{s_2}$, and
$\psi_2, \psi_3$ and $ \psi_4$ are active neutrinos
$\nu_{e}, \nu_{\mu}$ and $ \nu_{\tau}$.
In this identification the solar and atmospheric neutrino deficits 
are considered to be explained by
the small mixing MSW solution due to 
$\nu_e\rightarrow\nu_{s_1}$ and
the $\nu_\mu\leftrightarrow\nu_\tau$ oscillation in the vacuum, 
respectively.\footnote{
There is another possibility that the solar and atmospheric 
neutrino deficits are explained by the $\nu_e\leftrightarrow\nu_\mu$
and $\nu_\mu\leftrightarrow\nu_\tau$ oscillation, respectively.
 However, in such a case one light sterile neutrino plays no 
role in the oscillation phenomena and it can be reduced to the model 
considered in Ref. \cite{sue2}.} 
In Table 1 this identification has been assumed.   

Now we impose that the two flavors oscillation scheme is good enough for 
these oscillation processes.
And then we can determine some of the mixing parameters numerically
based on both oscillations in the $(\nu_{s_1}, \nu_e)$ sector with 
$\Delta M^2_{\bar 1\bar 2}$ and 
in the $(\nu_\mu, \nu_\tau)$ sector through 
a mode with $\alpha=\bar 3$ and $\beta=\bar 4$ which are shown in Table 1. 
As $U^{(l)}={\bf 1}$ is assumed here, the desired mixing 
angles in Eqs. (33) and
(34) can be realized by setting
\begin{equation}
16 ~{^<_\sim}~ {\mu_2 \over \mu_1} ~{^<_\sim}~ 32, \qquad\qquad
0.44 ~{^<_\sim}~ {\mu_3 \over \mu_4}~{^<_\sim}~2.3.
\end{equation}
In the following discussion we take these values as $\mu_2/\mu_1\simeq 25$ and 
$\mu_3/\mu_4\simeq 1$, for simplicity.
To investigate other oscillation processes it is convenient to use the
parameters $a$ and $b$ introduced in Eq. (26) which gives their physical
meanings as the ratio of coupling constants.
They can be also expressed as $\mu_2/\mu_3\equiv 10^{-a}$ and 
$\mu_4/\mu_5\equiv 10^{-b}$ and parametrize the mixing among 
different neutrino species.
There seem to be no quantitative constraints on  
these parameters at the present stage.
From the viewpoint that the solar and atmospheric neutrino deficits 
are explained by the two flavors scheme for the oscillation processes,
it is enough for them to be sufficiently large.

However, if we take account of the BBN constraint in the more quantitative way,
the $a$ and $b$ dependence of the mixing parameters seems to allow us 
to restrict the region of $a$ and $b$.
As mentioned earlier, the restriction on the number of the effective
neutrino species during the primordial nucleosynthesis gives the
condition on the $\nu_{s_2}$-$\nu_{e,\mu,\tau}$ sector.
Here it is sufficient to consider the most stringent one.
In the two flavors oscillation scheme 
when $m_{\nu_{s_2}}^2>m_{\nu_{e,\mu,\tau}}^2$ is satisfied, 
it can be formulated as \cite{nucl1,ssf},
\begin{equation}
\Delta m^2\sin^42\theta ~{^<_\sim}~3\times 10^{-6}~{\rm eV}^2 \quad
{\rm for}~~(\nu_{\mu,\tau},\nu_s).
\end{equation}
If we can apply this constraint to the $(\nu_\mu, \nu_{s_2})$ and 
$(\nu_\tau, \nu_{s_2})$ 
sectors in Table 1, nontrivial constraints on $b$ can be obtained.\footnote{
We should be careful in this application 
since this constraint has been derived in the two flavors oscillation scheme.
The BBN constraints are crucially affected by the interaction with the 
plasma at finite temperature so that the situation may be changed 
in many flavors case from the one of two flavor oscillation scheme.
Athough we need a numerical calculation for the correct analysis of 
this aspect, such a study is beyond the scope of this paper. 
However, this kind of consideration may be useful to show the
importance of the BBN constraint in the model building.}
This condition may be written as
\begin{equation}
10^{-4b}(M_{\bar 5}^2-M_{\bar 4}^2)~{^<_\sim}~10^{-6.7}.
\end{equation}
If we require that the heavier right-handed neutrino $\nu_{s_2}$
can be a dark matter, 
a value of $M_{\bar 5}$ should be fixed so as
to be a suitable value 
\footnote{This value should be changed dependeing on what kind of dark 
matter scenario is considered. Here we take a conservative value not
far from required values in the various models \cite{bes,dw,dark}.} 
and then the parameter $b$ should satisfy the following conditions:
\begin{equation}
b~{^>_\sim}~1.6~~(M_{\bar 5}\sim 1~{\rm eV}), \qquad
b~{^>_\sim}~2.1~~(M_{\bar 5}\sim 10~{\rm eV}), \qquad
b~{^>_\sim}~2.6~~(M_{\bar 5}\sim 10^2~{\rm eV}).
\end{equation}
If we consider the $(\nu_{s_1}, \nu_{e,\mu,\tau})$ 
sector with $m_{\nu_{s_1}}^2<m_{\nu_{e,\mu,\tau}}^2$,
through the BBN constraint given in Refs. \cite{nucl1,ssf}
there appears no condition on $a$ for the case of 
$\Delta M^2\sim 10^{-2.4}$
but $a~{^>_\sim}~1$ should be satisfied for the case of $\Delta M^2\sim 1$.
The situation completely depends on the suqared mass difference
among $\nu_{s_1}$ and $\nu_{e,\mu,\tau}$. Of course,
if we assume the existence of the large lepton asymmetry at the BBN
epoch, these constraints can disappear \cite{asym2}.
However, a lower bound on the parameter 
$a$ can be always obtained from the condition on the
amplitude in Eq. (7) such as $-4V_{\alpha f^\prime}V_{\alpha f}
V_{\beta f^\prime}V_{\beta f}~\le 1$. 
Using Table 1, a weak constraint on $a$ is brought as
$a~{^>_\sim}~0.15$ from this condition. 

In order to realize the desired squared 
mass differences in Eqs. (33) and (34), we need to require
\begin{equation}
M_{\bar 1}\sim 10^{-2.5}~{\rm eV}, \qquad 
\vert M_{\bar 3}^2-M_{\bar 4}^2\vert\sim 10^{-2.4}~{\rm
eV}^2.
\end{equation}
If we take $M_{\bar 5}\sim 10$~eV as an example and use 
the first one in Eq. (39), the 
parameters in our model defined by Eq. (8) should be settled by using
Eqs. (14) and (39) as,
\begin{eqnarray}
&&m_1\sim 10^{-5.4-(a+b)}M^{1\over 2}, \qquad\qquad m_2\sim
10^{-4-(a+b)}M^{1\over 2}, \nonumber \\
&&m_3 \sim m_4\sim 10^{-4-b}M^{1\over 2}, \qquad\qquad 
m_5\sim 10^{-4}M^{1\over 2},
\end{eqnarray} 
where we use a GeV unit, and $a$ and $b$ should satisfy 
the constraint given by Eq. (38) and $a~{^>_\sim}~0.15$. 
To make $\vert M_{\bar 3}^2-M_{\bar 4}^2\vert$ a value presented in Eq.
(39), there can be many possibilities for the values of
$M_{\bar 3}$ and $M_{\bar 4}$. Here we consider the following 
typical two cases :
$ {\rm (I)}~M_{\bar 3}<M_{\bar 4}\sim 10^{-1.2}~{\rm eV},$ and
${\rm (II)}~M_{\bar 3}\simeq M_{\bar 4}~(\gg 10^{-1.2}~{\rm eV}). $
In the case (II) we take two eigenvalues as 
$M_{\bar 3}\simeq M_{\bar 4}\sim 1$~eV which has been
studied in the various works \cite{pmodel} as an interesting 
example,
although such a choice requires a rather strict 
degeneracy between the third and fourth mass eigenvalues.
Numerical expressions of the 
oscillation parameters for each of these two cases
are given in the columns (I) and (II) of Table 1.
We should also note that these cases with certain values of $a$ and
$b$ can be realized as the two models
(I) and (II) presented in the previous section.

If we observe Table 1 taking account of Eq. (38) and $a~{^>_\sim}~0.15$, 
we immediately find that the very restricted oscillation modes can 
effectively 
occur and others are negligible because of the small amplitudes
( mixing angles ).
There are two $a$-independent processes
\begin{equation}
(\nu_{s_1}, \nu_e) ~~{\rm with}~~\Delta M_{\bar 1\bar 2}^2, 
\qquad (\nu_\mu, \nu_\tau) ~~{\rm with}~~\Delta M_{\bar 3\bar 4}^2, 
\end{equation}
and also as the $a$-dependent but
non-negligible interesting oscillation modes, we have 
\begin{equation}
(\nu_e, \nu_\mu) ~~{\rm with}~~\Delta M_{\bar 2\bar 3}^2, 
\Delta M_{\bar 2\bar 4}^2, 
\qquad (\nu_e, \nu_\tau) ~~{\rm with}~~\Delta M_{\bar 3\bar 4}^2.
\end{equation}
As already mentioned, two processes given in Eq. (41) can be treated 
within the two flavors oscillation 
scheme, as long as the parameter $a$ takes a suitable value which
can guarantee such a treatment.
In that case they can be used for the explanation of 
the solar and atmospheric neutrino problems in both cases of (I) and (II).
On the other hand, although some of the processes listed in Eq.  
(42) come out as the effects due
to many flavors existence, they may bring about the important
contributions to the $\psi_f\leftrightarrow\psi_{f^\prime}$ oscillation 
according to the value of $a$.
Next we examine both cases (I) and (II) in more detail and also 
discuss these processes in each case.

\subsection{Case I : $M_{\bar 3}< M_{\bar 4}\sim 10^{-1.2}~{\rm eV}$}
In this case \cite{sue1}, from Eqs. (13), (14), (39) and (40), we should
take the parameters as 
\begin{eqnarray}
&&{\cal A}\sim 10^{2(a+b)-0.7}M, \qquad {\cal B}\sim 
-10^{2(a+b)-3.5}M,\nonumber \\
&&{\cal C}\sim {\cal D}\sim {\cal E}\sim 10^{2b-2.5}M,\qquad 
{\cal F}\sim -10^{-2.2}M,
\end{eqnarray} 
where $M_{\bar 5}\sim 10$~eV and $a>0.5$ are assumed.
The allowed range of $a$ can determine the sign of ${\cal B}$, 
since ${\cal B}<0$ requires $({\cal B}-{\cal D})/{\cal D}<-1$ and 
then $a>0.5$. 
${\cal D}\not= {\cal E}$ should be reminded in order 
to resolve the mass degeneracy $M_{\bar 2}=M_{\bar 3}=0$.
The consistency of this model requires that the largest mass parameter 
${\cal A}$ should be smaller than $M_{\rm pl}$. This requirement brings about
an additional constraint on the parameters $a$ and $b$,
\begin{equation}
a+b ~{^<_\sim}~0.35 +\log_{10}\left({M_{\rm pl}\over M}\right)^{1\over 2}.
\end{equation} 
If we take $M\sim 10^{12}$~GeV which is realized in the examples
defined by Eq. (24) and use Eq. (38), we can constrain the value of
$a$, for example as follows: 
\begin{equation}
a~{^<_\sim}~1.75\hspace{2cm} {\rm for}~~ b~{^>_\sim}~2.1~~ (M_{\bar
5}\sim 10~{\rm eV}). 
\end{equation}
It should be also noted that
we can judge the phenomenological validity and consistency of the
constructed models based on the discrete symmetry 
such as the ones shown in Eq. (28) by
comparing it with Eq. (43) and studying whether $a$ and $b$ satisfy
the conditions given by Eqs. (37) and (44).
In the model defined by (I) of Eq. (28) we obtain $a\sim 1.5$ and
$b\sim 2.2$, which satisfy these conditions.

In the basis of these knowledge we can readily investigate the processes 
shown in Eq. (42).
As easily found in Table 1, if we take $a\sim 1.25$ and $M_{\bar 3}\sim
10^{-2.5}$, the oscillation
parameters of the first process with $\Delta M_{\bar 2\bar 3}^2$
 in Eq. (42) also seems to take 
appropriate values for 
the small mixing MSW solution of the solar neutrino problem.
In this case the light sterile neutrino may not be necessary for the
explanation of the solar neutrino deficit.
For smaller $a$, anyway, this process with larger 
$M_{\bar 3}$, $(\nu_e, \nu_\mu)$ with $\Delta M_{\bar 2\bar 4}^2$ 
and $(\nu_e, \nu_\tau)$ with $\Delta M_{\bar 3\bar 4}^2$ may be good
targets of long baseline experiments.
However, these processes in the interesting regions of $M_{\bar
3}$ and $a$
does not seem to be the ones suitable for the two flavors treatment.
Especially, in these processes the matter effect
seems not to be analytically estimated in the precise way and we may not 
be able to apply the condition given in Eq. (33) to this case naively.
The numerical analysis for the oscillations among 
many flavors will be indispensable for more detailed study.
If we want to guarantee the validity of the two flavor oscillation 
analysis for both the solar neutrino and atmospheric neutrino
problems due to $\nu_{s_1}\leftrightarrow \nu_e$ and
$\nu_\mu\leftrightarrow \nu_\tau$, we need to require 
$a~{^>_\sim}~1.3$ and then $M~{^<_\sim}~10^{13}$~GeV,
which can be satisfied in our model defined by Eq. (24).  
In such a case all other oscillation processes than 
$\nu_{s_1}\leftrightarrow\nu_e$ and $\nu_\mu\leftrightarrow\nu_\tau$ 
listed in Eq. (41) unfortunately seem to be inaccessible experimentally.

In this case all active neutrinos are too light to be a hot dark
matter but $\nu_{s_2}$ may be a warm dark matter with
$m_{\nu_{s_2}}=O(10-10^2)$ eV and $\Omega_{\nu_{s_2}}\sim 1$ \cite{dw}.
The main problem is how it is sufficiently produced at the early universe.
There seem to be two possibilities for its production.
If $\nu_{s_2}$ has an interaction with other light fields to
 be in the thermal equilibrium 
and then decouples as a relativistic particles, there is a relic \cite{hdform}
\begin{equation}
\Omega_{\nu_{s_2}}h^2={1\over g_*}~{m_{\nu_{s_2}} \over 8.5~ {\rm eV}},
\end{equation}  
where $h=H_0/(100{\rm km /sec/Mpc})$ and $H_0$ is the present Hubble
constant.
$g_*$ is the effective degrees of freedom of the light fields at the
$\nu_{s_2}$ decoupling time.
In the present model $\nu_{s_2}$ has the interaction with other fields 
due to the extra U(1) gauge symmetry which breaks down at a very high energy
scale $u_\ell$.  The above formula is applicable to $\nu_{s_2}$. 
If we assume $h\sim 0.5$ and $g_*~{^>_\sim}~300$ 
which is usually expected for this type of supersymmetric 
models at the decoupling
epoch in the very high energy scale,
we have $\Omega_{\nu_{s_2}}\sim 1$ for $M_{\bar 5}\sim 10^2$~eV.
If inflation occurs after the decoupling of $\nu_{s_2}$, however,
this possibility cannot be applyed.

Another possibility is the production through the
$\nu_{\mu,\tau}$-$\nu_{s_2}$ oscillation as suggested in Ref. \cite{dw}.
In this case $\nu_{\mu,\tau}$ are at first in the thermal equilibrium and then
decouples satisfying the same relation between $\Omega_\nu$ and $m_\nu$ 
given by Eq. (46). 
During this period $\nu_{s_2}$ is considered to be produced
from $\nu_{\mu,\tau}$ through the oscillation process
$\nu_{\mu,\tau}\leftrightarrow \nu_{s_2}$.
When we take this possibility, 
the constraint on $M_{\bar 5}$ comes from the requirement 
for both of the sufficient
abundance and the consistency with the BBN \cite{ssf}.
If we follow Ref. \cite{dw}, the ratio of the distribution
functions $f_s$ and $f_A$
of sterile and active neutrinos can be estimated in our model as
\begin{equation}
{f_s \over f_A} ={6.0 \over g_*^{1/2}}\left({M\mu_3\mu_5 \over {\rm
eV}}\right)^2\left({{\rm keV}\over M\mu_5^2}\right),\qquad 
{\Omega_{\nu_{s_2}}\over \Omega_A}={M_{\bar 5} f_s\over M_{\bar 4}f_A}.
\end{equation}  
By applying the hot relic relation given by Eq. (46) to these formulas
and remembering $M_{\bar 5}=M\mu_5^2$, 
we can derive the following relation:
\begin{equation}
M_{\bar 5}~{\mu_3\over \mu_5}=0.22h\Omega_{\nu_{s_2}}^{1/2}~{\rm eV}.
\end{equation}
On the other hand, the BBN constraint requires $f_s/f_A ~{^<_\sim}~
0.4$ and if we impose this on the latter of Eq. (47) and 
use Eq. (46),  we can obtain
\begin{equation}
M_{\bar 5} ~{^>_\sim}~230 h^2 \Omega_{\nu_{s_2}}~{\rm eV}.
\end{equation}
In this case we should take $g_*\sim 10.8$ and then
the required values for $M_{\bar 5}$ and
$b=\ln\left(\mu_5/\mu_3\right)$
are $M_{\bar 5}~{^>_\sim}~58$~eV and $b~{^>_\sim}~2.7$ 
for $\Omega_{\nu_{s_2}}\sim 1$ (WDM)
and $M_{\bar 5}~{^>_\sim}~17$~eV and 
$b~{^>_\sim}~2.4$ for $\Omega_{\nu_{s_2}}\sim 0.3$ (CHDM) \cite{sue2}.
This value of $M_{\bar 5}$ for the case of CDHM is so large that the
free streaming length of $\nu_{s_2}$ becomes too short and then seems not 
to contribute the structure formation at the supercluster scale \cite{phkc}.
Thus $\nu_{s_2}$ can be expected only to play the role as the warm dark
matter. In that case $a$ cannot take large value and $a~{^<_\sim}~1.1$.  

\subsection{Case II : $M_{\bar 3}\simeq M_{\bar 4}\sim 1$~eV}
In this case we need a rather strict fine tuning like
$M_{\bar 4}-M_{\bar 3}\sim10^{-2.7}$~eV. 
If we assume such a fine tuning, the parameters can be settled as 
\begin{eqnarray}
&&{\cal A}\sim 10^{2(a+b)-0.7}M, \qquad {\cal B}
\sim -10^{2(a+b)-3.5}M,\nonumber \\
&&{\cal C}\sim {\cal D} \sim 10^{2b-1}M,\qquad 
{\cal E}\sim 10^{2b-4}M, \qquad {\cal F}\sim -10^{-1}M
\end{eqnarray} 
where $M_{\bar 5}\sim 10$~eV and $a>1.3$ is assumed.
The condition given by Eq. (44) should be satisfied also in this case.
If we try to realize this case in terms of the model defined 
by (II) of Eq. (28), we obtain $a\sim 1.5$ and
$b\sim 2.2$ and the required conditions for $a$ and $b$ are fulfilled. 
The difference from the case (I) concerning the oscillation phenomena 
is that the rather large squared mass
difference such as $O(1)~{\rm eV}^2$ can appear between $\nu_e$ and
$\nu_{\mu,\tau}$. 

For such a squared mass difference, we can expect that the 
processes in Eq. (42) become 
interesting ones from the experimental viewpoint.
If we apply the results of BNL E776 \cite{bnl} and 
KARMEN \cite{karmen} experiments for the $\nu_e$ appearence through
$\nu_\mu\rightarrow \nu_e$ 
to the first process in Eq. (42), we can obtain a new lower bound on $a$.
Two flavors oscillation analysis of the data obtained by
these puts the most stringent bound on a mixing angle among $\nu_e$
and $\nu_\mu$ such as
$\sin^22\theta ~{^<_\sim}~7\times 10^{-3}$ for $\Delta
m^2=O(1)~{\rm eV^2}$.
Using Table 1, we can obtain $a~{^>_\sim}~1.23$ from this bound.
This also satisfies the constraint from the $\nu_e\rightarrow\nu_\mu$
transition from Burgey \cite{burgey} which requires $a~{^>_\sim}~1.0$
for the first one in Eq. (42) and also the constraint from the BBN
which was mentioned before.
   
In this context the interesting experimental results are the ones of LSND
\cite{lsnd}. The evidences for the oscillations
$\nu_\mu\rightarrow\nu_e$ and $\bar\nu_\mu \rightarrow\bar\nu_e$
have been reported there. 
One of nice features in this case is that these LSND results 
seem to be explained 
in the present model in terms of the first two processes in
Eq. (42).
In fact, as we take $\Delta M_{\bar 2\bar 3}^2, 
\Delta M_{\bar 2\bar 4}^2\sim 1~{\rm eV}^2$, 
for this squared mass difference the LSND
results require that the mixing angle should be $ 3\times 10^{-3}~{^<_\sim}~
\sin^22\theta~{^<_\sim}~ 1.5\times 10^{-2}$ for 
$\bar\nu_\mu\rightarrow\bar\nu_e$ and $1.5\times 10^{-3}~{^<_\sim}~
\sin^22\theta~{^<_\sim}~ 1.5\times 10^{-1}$ for $\nu_\mu\rightarrow\nu_e$.
These can be realized by taking $1.1~{^<_\sim}~a~{^<_\sim}~1.4$ and
$0.6~{^<_\sim}~a~{^<_\sim}~1.6$, respectively.
If these are combined with the above results of BNL E776 and KARMEN, 
$a$ can be constrained to a very narrow region 
$1.23~{^<_\sim}~a~{^<_\sim}~1.4$. 
If $a$ takes a small value which cannot explain the LSND results, the 
second process in Eq. (42) may become a very interesting target in the
future experiments.
An interesting aspect of this case is that the above region of $a$ relevant 
to the LSND results can put the constraint on the value of $M_{\bar 5}$ which 
is the mass of a candidate of the dark matter.
In fact, if we assume $a\sim 1.4$ and $M\sim 10^{12}$~GeV as an example, 
we obtain the bound on $M_{\bar 5}$ as $M_{\bar 5}~
{^<_\sim}~36$~eV by combining
Eq. (44) and the BBN constraint given by Eq. (37).

Anyway this case in our framework corresponds to an interesting
realization of the model
which has been pointed out by various authors \cite{pmodel}, in 
which the solar and atmospheric neutrino deficits and 
the LSND results can be simultaneously 
explained and additionally $\nu_\mu$ and $\mu_\tau$ can be the
hot dark matter candidates in the CHDM scenario.
In this case we may naturally ask the physical role of $\nu_{s_2}$.
The first interest is whether $\nu_{s_2}$ can have some affection 
for the structure formation or not.
We can estimate this by using Eqs. (48) and (49) as in the previous
case. For example, if we assume $\Omega_{\nu_{s_2}}\sim 0.1$, we have
$M_{\bar 5}~{^>_\sim}~5.8$~eV and also $b~{^>_\sim}~2.2$ which is included in 
the allowed region presented by Eq. (37).
This suggests that $\nu_{s_2}$ may have some affection on the structure
formation as a part of hot dark matter in the CHDM scenario.
Another interesting possibility of the physical role of $\nu_{s_2}$ 
may be an effect on the leptogenesis discussed in Ref. \cite{ster2}.
We do not study it here but it may be an interesting aspect of our model.

\subsection{Relation to the charged lepton sector}
It may be useful to comment on the constraint on the charged lepton 
mass matrix in the present framework.
Although we have assumed that the charged lepton mass matrix is diagonal up 
to now, it is also related to the neutrino 
oscillation phenomena through the mixing matrix $V$ as shown in
Eq. (7).
If we once fix the mixing matrix of the neutrino sector as Eq. (11), 
the neutrino oscillation data can constrain 
the structure of the charged lepton mass matrix. 

Here we consider two typical examples as the charged 
lepton mass matrix.
The first example (A) is represented by Eq. (30) which is obtained in
the basis of the discrete symmetry.
Using the mixing matrix $U^{(l)}$ given in Eq. (32), 
we can obtain the mixing matrix elements 
$V_{\alpha f}$ defined in section 2 as follows,
\begin{eqnarray}
&&V_{\bar 2 e}\simeq 1,\quad
V_{\bar 2 \mu}\simeq -e^{i\sigma}{\mu_2 \over \mu_3},\quad
V_{\bar 2 \tau}\simeq 
-e^{i\sigma}{\mu_2 \over \mu_3}{k_E^{\mu\tau}\over k_E^{\tau\tau}},\nonumber \\
&&V_{\bar 3 e}\simeq {1\over\sqrt 2}{\mu_2 \over \mu_3},\quad
V_{\bar 3 \mu}\simeq {e^{i\sigma} \over \sqrt 2},\quad
V_{\bar 3\tau}\simeq -{e^{i\tau} \over \sqrt 2}, \nonumber \\
&&V_{\bar 4 e}\simeq {1\over\sqrt 2}{\mu_2 \over \mu_3},\quad
V_{\bar 4 \mu}\simeq {e^{i\sigma} \over \sqrt 2},\quad
V_{\bar 4 \tau}\simeq {e^{i\tau} \over \sqrt 2}, 
\end{eqnarray}
where we have introduced phases $\sigma$ and $\tau$ for completeness.

As an another typical phenomenological example (B),
we adopt a Fritzsch mass matrix \cite{frit} for the charged lepton sector.  
Although it is not relevant to our construction of the neutrino sector 
based on the discrete symmetry given in Table 2, it will be useful to
find the feature of the mixing matrix presented in Eq. (11).
Using a well-known formula in the diagonalization of the Fritzsch mass 
matrix as $U^{(l)}$, we can get the mixing matrix elements 
for the lepton sector as\footnote{
These expressions are somehow different from the ones given in Ref.
\cite{sue1} since we take the mass hierarchy given by Eq. (9) which should be
assumed for the explanation of the solar neutrino problem.}
\begin{eqnarray}
&&V_{\bar 2 e}\simeq 1,\quad
V_{\bar 2 \mu}\simeq -\sqrt{m_e \over m_\mu}
 -e^{i\sigma}{\mu_2 \over \mu_3},\quad
V_{\bar 2 \tau}\simeq 
-e^{i\sigma}{\mu_2 \over \mu_3}\sqrt{m_\mu \over m_\tau},\nonumber \\
&&V_{\bar 3 \mu}\simeq {1 \over \sqrt 2}\left(e^{i\sigma}
+e^{i\tau}\sqrt{m_\mu\over m_\tau}\right)  ,\quad
V_{\bar 3 \tau}\simeq {1 \over \sqrt 2}
\left(e^{i\sigma}\sqrt{m_\mu\over m_\tau}-e^{i\tau}\right),\nonumber \\
&&V_{\bar 4\mu}\simeq {1 \over \sqrt 2}
\left(e^{i\sigma}-e^{i\tau}\sqrt{m_\mu\over m_\tau}\right),\quad
V_{\bar 4\tau}\simeq {1 \over \sqrt 2}\left(e^{i\sigma}
\sqrt{m_\mu\over m_\tau}+e^{i\tau}\right), \\
&&V_{\bar 3 e}\simeq {1\over\sqrt 2}{\mu_2 \over \mu_3}
+{e^{i\sigma}\over \sqrt 2}\sqrt{m_e \over m_\mu}
+{e^{i\tau} \over \sqrt 2}\sqrt{m_e \over m_\tau},\quad
V_{\bar 4 e}\simeq 
{1\over\sqrt 2}{\mu_2 \over \mu_3}
+{e^{i\sigma}\over\sqrt 2}\sqrt{m_e \over m_\mu}
-{e^{i\tau} \over \sqrt 2}\sqrt{m_e \over m_\tau},\nonumber  
\end{eqnarray}
where $m_e, m_\mu$ and $m_\tau$ are charged lepton mass eigenvalues.

In both cases $V$ is found to have the similar form except that
the latter example has extra contributions to the off-diagonal part
compared to the formar one.
We present numerical values of the representative 
oscillation parameters for suitable settings of $a$ and $b$ in Table 3.
From this table we find that both solar and atmospheric neutrino deficits 
can be simultaneously explained for these charged lepton mass matrices. 
Related to this point it may be useful to note that these charged 
lepton mass matrices have no effect on $V_{\bar 1 e}$ and $V_{\bar 2s_1}$, 
which are the elements of the above mixing matrix relevant to 
the $\nu_e\leftrightarrow\nu_{s_1}$ oscillation as found from Eq. (7).
This feature is very different from the models in which
the similar mass matrices are assumed for both the charged lepton and 
neutrino sectors. 
In those models,  
the mixings $V_{\bar 2\mu}$ and $V_{\bar 3e}$ 
has a tendency to become too large 
due to the contribution from the charged
lepton sector without assuming suitable values of phases
to explain the solar neutrino deficit by the small mixing 
MSW solution for $\nu_e\rightarrow\nu_\mu$ 
if we keep $V_{\bar 3 \tau}$ and $V_{\bar 4 \mu}$ 
to be suitable for the explanation 
of the atmospheric $\nu_\mu$ deficit due to $\nu_\mu
\rightarrow\nu_\tau$ \cite{bs,matrix2}.
This aspect also appears in the $(\nu_e, \nu_\mu)$ mixing of the case 
(B) in Table 3.
The present model does not suffer from this problem 
as a direct result that the solar neutrino deficit is explained 
by the $\nu_e\leftrightarrow\nu_{s_1}$ oscillation due to the 
introduction of a sterile neutrino.
These two examples show that as long as the $U^{(l)}$ is 
approximately diagonal in the similar way to these examples, our scenario is 
always expected to be applicable independently of the details of the charged 
lepton mass matrix. 

When we adopt these charged lepton mass matrices,
the LSND results can be also 
explained in the case (II) in the same way as discussed in the
previous part as long as we take $a$ in a suitable region around
$a\sim 1.3$.
The situation on the consistency with the BBN constraint is 
also similar to the case of $U^{(\ell)}=1$ since the charged lepton
sector has no important effect in the $(\nu_{s_2}, \nu_\tau)$ sector.
No contradiction happens against the BBN constraint
through the oscillations in the $(\nu_{s_2}, \nu_\tau)$ sector
if $b$ and $M_{\bar 5}$ are in the suitable region shown in Eq. (38).
This feature can be seen in Table 3. For other processes the same discussions 
presented in the present section are also valid in the present case.

\begin{figure}[hbt]
\begin{center}
\begin{tabular}{c|cccl}
$(\psi_f,~\psi_{f^\prime})$ & $\Delta M^2_{\alpha\beta}$ 
& \multicolumn{2}{c}{$-4V_{\alpha f^\prime}V_{\alpha f}
V_{\beta f^\prime}V_{\beta f}$}& Relevant Phenomena\\ 
&& (A) & (B) & \\ \hline\hline
$(\nu_{s_1}, \nu_e)$ & $\Delta M_{\bar 1\bar 2}^2$ &$6.4\times 10^{-3}$ 
&$6.4\times 10^{-3}$ & Solar Neutrino Deficit\\
$(\nu_\mu, \nu_\tau)$ & $\Delta M_{\bar 3\bar 4}^2$ &1& 0.9 
& Atmosphecic Neutrino Deficit\\ 
$(\nu_e, \nu_\mu)$ & $\Delta M_{\bar 2\bar 3}^2$ & $5.1\times 10^{-3}$ 
&$4.1\times 10^{-2}$ & LSND\\ 
$(\nu_e, \nu_\mu)$ & $\Delta M_{\bar 2\bar 4}^2$ & $5.1\times 10^{-3}$ 
&$1.9\times 10^{-2}$ & LSND\\ 
$(\nu_e, \nu_\tau)$&$\Delta M_{\bar 3\bar 4}^2$ &$2.6\times 10^{-3}$ 
&$1.3\times 10^{-2}$ &  Prediction of the model\\
$(\nu_e, \nu_\tau)$ & $\Delta M_{\bar 2\bar 4}^2$ &$5.1\times 10^{-4}$ 
& $3.1\times 10^{-3}$ & Prediction of the model\\
$(\nu_\mu, \nu_\tau)$ & $\Delta M_{\bar 2\bar 3}^2$ &$5.0\times 10^{-4}$ 
& $2.7\times 10^{-3}$ & Prediction of the model\\
$(\nu_{s_1}, \nu_\mu)$ & $\Delta M_{\bar 2\bar 3}^2$&$8.0\times 10^{-6}$ 
& $2.4\times 10^{-5}$ & BBN Constraint\\
$(\nu_{s_1}, \nu_\tau)$ & $\Delta M_{\bar 2\bar 4}^2$&$8.0\times 10^{-6}$ 
& $2.4\times 10^{-6}$ & 
BBN Constraint\\
$(\nu_{s_2}, \nu_e)$ & $\Delta M_{\bar 4\bar 5}^2$&$1.6\times 10^{-7}$ 
& $4.1\times 10^{-8}$ & 
BBN Constraint\\
$(\nu_{s_2}, \nu_\mu)$ & $\Delta M_{\bar 4\bar 5}^2$&$6.4\times 10^{-5}$ 
& $2.3\times 10^{-4}$ & 
BBN Constraint\\
$(\nu_{s_2}, \nu_\tau)$ & $\Delta M_{\bar 4\bar 5}^2$&$6.4\times 10^{-5}$ 
& $6.0\times 10^{-4}$ & 
BBN Constraint

\end{tabular}\vspace{4mm}\\
\end{center}
{\bf Table 3}~~Numerical values of neutrino oscillation parameters 
for the charged lepton mass matrices (A) and (B).
We take parameters as $a\sim 1.3$,  $b\sim 2.4$, $\sigma\sim\tau\sim
0$ and $k_E^{\mu\tau}/k_E^{\tau\tau}=0.1$.
Assuming the mass differences which can explain the solar and
atmospheric neutrino deficits, only the dominant contributions 
in each $(\psi_f,~\psi_{f^\prime})$ sector are presented. 
\end{figure}
 
\section{Summary}
We proposed the neutrino mass matrix 
in the $3\nu_L+3\nu_R$ framework, which could be constructed using the 
nonrenormalizable terms in the superpotential constrained by the
suitable discrete symmetry.
We showed that it
could explain the solar and atmospheric neutrino deficits and
 give a dark matter candidate.
We also discussed that there could be two typical parameter 
settings for Yukawa coupling constants, which brought
about rather different phenomenological features.
An interesing aspect of this model is that one of these parameter
settings can also realize the mass and mixing pattern which has been
known to explain the LSND results, simultaneously.
It may be considered as another interesting feature of our mass matrix 
that it can explain both deficits of the solar and atmospheric neutrinos
without severely constraining the charged lepton mass matrix as long
as it has no large off-diagonal elements.
Although these features simply come from the extension of the
parameter space due to the introduction of the new light
sterile neutrino species, this kind of investigation can be considered 
to have a sufficient meaning to show the way for the extension of 
the neutrino sector.

The introduction of the light sterile neutrinos is usually
considered to be artificial. However,  
their appearence seems to be not so unnatural as shown in this paper
if we take account of the generation structure of quarks and charged leptons
and also assume the constrained nonrenormalizable superpotential.  
One of such simple and promising candidates may be the extra U(1)
models coming from the $E_6$ models inspired by the perturbative 
superstring, in which the group theoretical constraints on 
the Yukawa couplings are very weak.
In such a case all but one or two of the right-handed neutrinos 
can be generally very light by the cooperation of both 
the superpotential constrained by the discrete symmetry 
and the extra U(1) D-flat direction. 
They can play the important role in the neutrino physics such as neutrino 
oscillations.
Although this scheme seems to be successful,
it is generally not so easy to yield small neutrino masses and 
to induce the neutrino oscillations without bringing other phenomenological
difficulties like proton decay in this framework \cite{esix,s}.
The simultaneous explanation of them will be the next step to build
the realistic models in this direction.
Anyway we believe that it will be worthy to proceed the 
further investigation of this kind of possibilities.

\vspace{7mm}

{\large\bf Acknowledgments}

The author would like to thank T.~Goldman for calling his attention to
the LSND results.
This work has been supported
by a Grant-in-Aid for Scientific Research (C) from the Ministry of Education, 
Science and Culture(\#08640362).

\newpage 

\end{document}